\documentclass[aps,prc,floatfix,twocolumn,showpacs,amsmath]{revtex4}
\usepackage{dcolumn}
\usepackage{mathrsfs}
\usepackage{supertabular}
\usepackage{color,graphicx}
\usepackage[bookmarksopen]{hyperref}
\newcommand{\bm}{\bibitem}

\def  \a    {\alpha}

\def  \g    {\gamma}
\def  \G    {\Gamma}
\def  \l    {\lambda}
\def  \L    {\Lambda}
\def  \o    {\omega}

\def  \s    {\sigma}

\def  \p    {\pi}
\def  \P    {\Pi}
\def  \m    {\mu}
\def  \n    {\nu}

\def  \t    {\tau}

\def  \f    {\frac}
\def  \lt   {\left}
\def  \rt   {\right}
\def  \th   {\theta}

\def  \sq   {\sqrt}
\def  \ra   {\rightarrow}

\def  \dt   {\delta}
\def  \Dt   {\Delta}
\def  \ep   {\epsilon}
\def  \z    {\zeta}

\def  \be   {\begin{equation}}
\def  \ee   {\end{equation}}
\def  \ba   {\begin{array}}
\def  \ea   {\end{array}}
\def  \bea  {\begin{eqnarray}}
\def  \eea  {\end{eqnarray}}
\def  \nn   {\nonumber}
\def  \bd   {\begin{displaymath}}
\def  \ed   {\end{displaymath}}
\def  \bse  {\begin{subequations}}
\def  \ese  {\end{subequations}}
\def  \bwt  {\begin{widetext}}
\def  \ewt  {\end{widetext}}

\def  \vp   {{\bf p}}
\def  \vs   {{\bf \sigma}}

\def  \rw   {\rho\omega}
\def  \grw  {g_\rho g_\omega}
\def  \xp   {x^\prime}
\def  \kp   {k^\prime}

\def  \vk   {{\bf k}}
\def  \vkp  {{\bf k}^\prime}
\def  \tp   {{t^\prime}}
\def  \sp   {{s^\prime}}
\def  \bra  {\mathinner{\langle}}
\def  \ket  {\mathinner{\rangle}}

\begin{document}
\title{Matter induced charge symmetry violating NN potential}
\author{Subhrajyoti Biswas}
\author{Pradip Roy}
\author{Abhee K. Dutt-Mazumder}
\address{Saha Institute of Nuclear Physics,
1/AF Bidhannagar, Kolkata-700 064, INDIA}

\medskip
\date{\today}
\begin{abstract}

We construct density dependent Class $III$ charge symmetry violating (CSV)
potential due to mixing of $\rho$-$\omega$ meson with off-shell corrections.
Here in addition to the usual vacuum contribution, the matter induced mixing
of $\rho$-$\omega$ is also included. It is observed that the contribution of
density dependent CSV potential is comparable to that of the vacuum contribution.

\end{abstract}
\vspace{0.08 cm}
\pacs{21.65.Cd, 13.75.Cs, 13.75.Gx, 21.30.Fe}
\maketitle

\section{introduction}

The exploration of symmetries and their breaking have always
been an active and interesting area of research in nuclear physics.
One of the well known examples, that can be cited here, is the
nuclear $\beta$ decay which violates parity that led to the
discovery of the weak interaction. Our present concern, however,
is the strong interaction where, in particular, we focus attention
to the charge symmetry violation (CSV) in nucleon-nucleon ($NN$)
interaction.

Charge symmetry implies invariance of the $NN$ interaction under
rotation in isospin space, which in nature, is violated. The
CSV, at the fundamental level is caused by the finite mass difference
between up $(u)$ and down $(d)$ quarks \cite{Nolen69,Henley69,Henley79,
Miller90,Machleidt89,Miller95}. As a consequence, at the hadronic level,
charge symmetry (CS) is violated due to non-degenerate mass of hadrons
of the same isospin multiplet. The general goal of the research in this
area is to find small but observable effects of CSV which might provide
significant insight into the strong interaction dynamics.

There are several experimental data which indicate CSV in $NN$ interaction.
For instance, the difference between $pp$ and $nn$ scattering lengths
at $^1$S$_0$ state is non-zero \cite{Miller90,Howell98,Gonzalez99}.
Other convincing evidence of CSV comes from the binding energy difference
of mirror nuclei which is known as Okamoto-Nolen-Schifer (ONS) anomaly
\cite{Nolen73,Okamoto64,Garcia92}. The modern manifestation
of CSV includes difference of neutron-proton form factors,
hadronic correction to $g-2$ \cite{Miller06}, the observation
of the decay of $\Psi^{\prime} (3686) \ra (J/\Psi) \p^0 $ etc
\cite{Miller06}.

In nuclear physics, one constructs CSV potential to see its
consequences on various observables. The construction of CSV
potential involves evaluation of the $NN$ scattering diagrams
with intermediate states that include mixing of various isospin
states like $\rho$-$\omega$ or $\pi$-$\eta$ mesons.
The former is found to be most dominant \cite{Coon87,Henley79,McNamee75,
Coon77,Blunden87,Sidney87} which we consider here.

Most of the calculations performed initially to construct CSV
potential considered the on-shell \cite{Blunden87} or constant
$\rho$-$\omega$ mixing amplitude \cite{Sidney87}, which are
claimed to be successful in explaining various CSV observables
\cite{Sidney87,Machleidt01}. This success has been
called into question \cite{Cohen95,Piekarewicz92} on the ground
of the use of on-shell mixing amplitude for the construction
of CSV potential. First in \cite{Piekarewicz92} and then in
\cite{Goldman92,Krein93,Connell94,Coon97,Hatsuda94}, it is shown
that the $\rho$-$\omega$ mixing has strong momentum dependence
which even changes its sign as one moves away from the $\rho$
(or $\omega$) pole to the space-like region which is relevant
for the construction of the CSV potential. Therefore inclusion
of off-shell corrections are necessary for the calculation of
CSV potential. We here deal with such mixing amplitude induced
by the $N$-$N$ loop incorporating off-shell corrections.

In vacuum, the charge symmetry is broken explicitly due to the
non-degenerate nucleon masses. In matter, there can be another source
of symmetry breaking if the  ground state contains unequal number of
neutrons ($n$) and protons ($p$) giving rise to ground state induced
mixing of various charged states like $\rho$-$\omega$ meson even
in the limit $M_n=M_p$. This additional source of symmetry
breaking for the construction of CSV potential has, to the best of our
knowledge, not been considered before.

The possibility of such matter induced mixing was first studied in
\cite{Abhee97} and was subsequently studied in \cite{Broniowski98,
Abhee01,Kampfer04,Roy08}. For the case of $\pi$-$\eta$ meson also such
asymmetry driven mixing is studied in \cite{Biswas06}. But none of these
deal with the construction of two-body potential and the calculations
are mostly confined to the time-like region where the main motivation
is to investigate the role of such matter induced mixing on the
dilepton spectrum observed in heavy ion collisions, pion form factor,
meson dispersion relations etc. \cite{Broniowski98,Roy08}.
In Ref.\cite{Saito03}, attempt has been made to calculate the
density dependent CSV potential where only the effect of the scalar mean
field on the nucleon mass is considered excluding the possibility of matter
driven mixing. All existing matter induced mixing calculations, however,
suggest that, at least in the $\rho$-$\omega$ sector, the inclusion
of such a matter induced mixing amplitude into the two body $NN$ interaction
potential can significantly change the results both qualitatively and
quantitatively. It is also to be noted that such mixing amplitudes,
in asymmetric nuclear matter (ANM), have non-zero contribution even if
the quark or nucleon masses are taken to be equal \cite{Abhee97,Broniowski98,
Abhee01,Kampfer04,Roy08}. We consider both of these mechanisms to
construct the CSV potential.

\begin{figure}[htb]
\begin{center}
\resizebox{6.5cm}{2.5cm}{\includegraphics[]{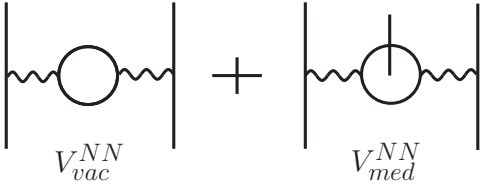}}
\caption{Feynman Diagrams that contribute to the construction of CSV
NN potential in matter. Solid lines represent nucleons and
wavy lines stand for vector mesons. \label{fig00}}
\end{center}
\end{figure}

Physically, in dense system, intermediate mesons might be
absorbed and re-emitted from the Fermi spheres. In symmetric
nuclear matter (SNM) the emission and absorption  involving
different isospin states like $\rho$ and $\omega$ cancel when
the contributions of both the proton and neutron Fermi spheres
are added provided the nucleon masses are taken to be equal.
In ANM, on the other hand, the unbalanced contributions coming
from the scattering of neutron and proton Fermi spheres, lead to
the mixing which depends both on the density $(\rho_B)$ and the
asymmetry parameter $[\a = (\rho_n - \rho_p)/\rho_B] $. Inclusion
of this process is depicted by the second diagram in Fig.\ref{fig00}
represented by $V^{NN}_{med}$ which is non-zero even
in symmetric nuclear matter if explicit mass differences of nucleons
are retained. In the first diagram, $V^{NN}_{vac}$ involves NN loop
denoted by the circle. The other important element which we include
here is the contribution coming from the external legs. This
is another source of explicit symmetry violation which significantly
modify the CSV potential in vacuum as has been shown only recently by
the present authors \cite{Biswas08}.

This paper is organized as follows. In Sec.II we present the formalism
where the three momentum dependent $\rho^0$-$\omega$ mixing amplitude
is calculated to construct the CSV potential in matter. The numerical results
are discussed in Sec.III. Finally, we summarize in Sec.IV.

\section{formalism}

We start with the following effective Lagrangians to describe $\o NN$
and $\rho NN$ interactions:

\bse
\bea
\mathcal{L}_{\o NN} & = &  {\rm g_\o}\bar{\Psi}\g_\m\Phi^\m_\o \Psi,
\label{lag:omega}\\
\mathcal{L}_{\rho NN} & = & {\rm g_\rho} \bar{\Psi}\lt[\g_\m +\f{C_\rho}{2M}
\sigma_{\m\n}\partial^\m \rt]\vec{\t} \cdot {\bf \Phi}^\n_\rho \Psi,
\label{lag:rho}
\eea
\ese

where $C_{\rho} = f_{\rho}/{\rm g_{\rho}}$ is the ratio of vector
to tensor coupling, $M$ is the average nucleon mass and $\vec{\t}$ is
the isospin operator. $\Psi$ and $\Phi$ represent the nucleon and meson
fields, respectively, and ${\rm g}$'s stand for the meson-nucleon coupling
constants. The tensor coupling of $\omega$ is not included in the present
calculation as it is negligible compared to the vector coupling.

The matrix element, which is required for the construction
of CSV $NN$ potential is obtained from the relevant Feynman
diagram \cite{Biswas08}:

\nointerlineskip
\bea
\mathcal{M}^{NN}_{\rho \omega}(q)\!\! &=& \!\! [\bar{u}_N(p_3)\G_\rho^\mu(q) u_N(p_1)]~
\Dt_{\mu \alpha}^\rho(q) \Pi^{\alpha \beta}_{\rho \omega}(q^2) \nn \\
&\times & \!\! \Dt_{\beta \nu}^{\omega}(q)~
[\bar{u}_N(p_4)\G_{\omega}^{\nu}(-q) u_N(p_2)]. \label{ma0}
\eea

In the limit $q_0 \ra 0$, Eq.(\ref{ma0}) gives the momentum space CSV $NN$
potential, $V^{NN}_{CSV}({\bf q})$. Here $\G^\m_\o(q) = {\rm g_\o} \g^\m $,
$\G^\n_\rho (q) = {\rm g_\rho} \lt[\g^\n - \f{C_\rho}{2M}i\s^{\n\l}q_\l\rt]$
denote the vertex factors, $u_N$ is the Dirac spinor and $\Dt^i_{\m\n}(q)$,
$(i = \rho, \omega)$ is the meson propagator. $p_j$ and $q$ are the four momenta
of nucleon and meson, respectively.

In the present calculation, $\rho$-$\omega$ mixing amplitude ({\em i.e.}
polarization tensor) $\Pi^{\m\n}_{\rho\omega}(q^2)$ is generated
by the difference between proton and neutron loop contributions:

\bea
\Pi^{\mu \nu}_{\rw}(q^2) = \Pi^{\mu \nu (p)}_{\rw}(q^2)
-\Pi^{\mu \nu (n)}_{\rw}(q^2).
\label{se}
\eea

\noindent Explicitly, the polarization tensor is given by

\nointerlineskip
\bea
&\!\!&i\Pi^{\m\n (N)}_{\rho\omega}(q^2)\!\! =\!\!\!\int \!\! \f{d^4k}{(2\pi)^4}
{\rm Tr}\!\lt[\!\G^\m_\o (q) G_N(k) \G^\n_\rho(-q) G_N(k\!+\!q)\!\rt],\nn\\
&\!\!&\label{pol:total0}
\eea

\noindent where $k=(k_0,{\bf k})$ denotes the four momentum of the
nucleon in the loops ({\em see} Fig.\ref{fig00}) and $G_N$ is the in-medium
nucleon propagator consisting of free $(G^F_N)$ and density dependent $(G^D_N)$
parts \cite{Serot86},

\bse
\label{nucl:prop}
\bea
G^F_N(k)\!\!&=&\!\!\frac{k\!\!\!/+M_N}{k^2-M^2_N+i\z}, \label{nucl:prop_vac} \\
G^D_N(k)\!\!&=&\!\!\frac{i\pi}{E_N}(k\!\!\!/+M_N)\delta(k_0-E_N)\theta(k_N-|{\bf k}|)
\label{nucl:prop_med}
\eea
\ese

\noindent The subscript $N$ stands for nucleon index ({\em i.e.} $N=p$ or $n$),
$k_N$ denotes the Fermi momentum of nucleon, nucleon energy $E_N=\sqrt{M^2_N+k^2_N}$
and nucleon mass is denoted by $M_N$. $\theta(k_N-|{\bf k}|)$ is the Fermi distribution
function at zero temperature.

The origin of $G^D_N(k)$, in addition to the free propagator resides in the fact that
here one deals with vacuum containing real particles which when acted upon the annihilation
operator does not vanish ({\em see } {\bf Appendix A} for details). The appearance
of the delta function in Eq.(\ref{nucl:prop_med}) indicates the nucleons are on-shell while
$\theta(k_N-|{\bf k}|)$ ensures that propagating nucleons have momentum less than $k_N$.

Likewise, the polarization tensor of Eq.(\ref{pol:total0}) also contains
a vacuum $[\Pi^{\m\n (N)}_{vac}(q^2)]$ and a density dependent
$[\Pi^{\m\n (N)}_{med}(q^2)]$ parts as shown in Fig.\ref{fig00}.
It is to be noted that the density dependent part given by the combination
of $G^F_NG^D_N+G^D_NG^F_N$ corresponds to scattering that we have discussed
already, whereas the term proportional to $G^D_NG^D_N$ vanishes for low energy
excitation \cite{Chin77}. The vacuum part, {\em viz.}
$[\Pi^{\m\n (N)}_{vac}(q^2)]$ on the other hand involves $G_N^FG_N^F$ which
gives rise to usual CSV part of the potential due to the splitting of
the neutron and proton mass.

It might be worthwhile to mention here that Eq.(\ref{nucl:prop_med}) can
induce charge symmetry breaking in asymmetric nuclear matter due to
the appearance of the Fermi distribution function in the propagator
itself which can distinguish between neutron and proton, even if
their mass are taken to be degenerate. Evidently, this is an exclusive
medium driven effect where, as mentioned already in the introduction,
the charge symmetry is broken by the ground state. The total charge
symmetry breaking would involve both the contributions where, it is
clear that even for $\alpha=0$, the medium dependent term can contribute
if the non-degenerate nucleon masses are considered.

Note that the polarization tensor $\Pi^{\m\n}_{\rho\omega}(q^2)$ can
be expressed as the sum of longitudinal component $[\Pi^L_{\rho\omega}(q^2)]$
and transverse component $[\Pi^T_{\rho\omega}(q^2)]$ which will be useful to
simplify the matrix element given in Eq.(\ref{ma0}).

\nointerlineskip
\bea
\Pi^{\m\n}_{\rho\omega}(q^2) = \Pi^L_{\rho\omega}(q^2) A^{\m\n} +
\Pi^T_{\rho\omega}(q^2) B^{\m\n},
\eea

\noindent where $A^{\m\n}$ and $B^{\m\n}$ are the longitudinal and transverse
projection operators \cite{Kapusta}. We define $\Pi^L_{\rw}=-\Pi^{00}_{\rw}
+ \Pi^{33}_{\rw}$ and $\Pi^T_{\rw}=\Pi^{11}_{\rw}=\Pi^{22}_{\rw}$.

We, in the present calculation, use the average of longitudinal and transverse
components of the polarization tensor instead of $\Pi^L_{\rw}$ and $\Pi^T_{\rw}$.
The average mixing amplitude is denoted by

\nointerlineskip
\bea
\bar{\Pi}(q^2) &=& \f{1}{3}\lt[\Pi^L_{\rw}(q^2)+2\Pi^T_{\rw}(q^2) \rt] \nn \\
               &=& \bar{\Pi}_{vac}(q^2) + \bar{\Pi}_{med}(q^2).
\label{pol:av0}
\eea

In the last line of Eq.(\ref{pol:av0}), $\bar{\Pi}_{vac}(q^2)$ and
$\bar{\Pi}_{med}(q^2)$ denote the average mixing amplitudes of vacuum
and density dependent parts, respectively.

To obtain $\bar{\Pi}_{vac}(q^2)$ and $\bar{\Pi}_{med}(q^2)$ one would calculate
the total polarization tensor given in Eq.(\ref{pol:total0}). After evaluating
the trace of Eq.(\ref{pol:total0}), we find the following vacuum and density
dependent parts of the polarization tensor.

\bea
\Pi^{\m\n(N)}_{vac}(q^2) &=& Q^{\m\n}\lt[\Pi^{vv(N)}_{vac}(q^2)
+ \Pi^{tv(N)}_{vac}(q^2) \rt], \label{pol:vac0}
\eea

\noindent and

\bea
\Pi^{\m\n\{vv(N)\}}_{med}(q^2) &=& 16{\rm \grw} \int\f{d^3k}{(2\pi)^32E_N}
\theta(k_N-|{\bf k}|) \nn \\
&\times& \lt[\f{q^2K^{\m\n} -(q\cdot k)^2Q^{\m\n}}{q^4 - 4(q\cdot k)^2} \rt],
\label{pol:dens_vv0} \\
\Pi^{\m\n\{tv(N)\}}_{med}(q^2) &=& 4{\rm \grw} C_\rho \int \f{d^3k}{(2\pi)^32E_N}
\theta(k_N-|{\bf k}|) \nn \\
&\times& \lt[\f{q^4Q^{\m\n}}{q^4-4(q\cdot k)^2} \rt],
\label{pol:dens_tv0}
\eea

\noindent where $Q^{\m\n} = (-g^{\m\n} + q^\m q^\n/q^2)$ and
$K^{\m\n} = \lt(k^\m -(q\cdot k)\f{q^\m}{q^2}\rt)\lt(k^\n -(q\cdot k)\f{q^\n}{q^2}\rt)$.
It is to be mentioned that both $\Pi^{\m\n}_{vac}(q^2)$ and $\Pi^{\m\n}_{med}(q^2)$
obey the current conservation as $q_\m Q^{\m\n} = q_\n Q^{\m\n} = 0$ and
$q_\m K^{\m\n} = q_\n K^{\m\n} = 0$. The superscripts $vv$ and $tv$ in Eqs.(\ref{pol:vac0})
-(\ref{pol:dens_tv0}) indicate the vector-vector and tensor-vector interactions, respectively.
The dimensional counting shows that vacuum part of the polarization tensor
[both $\Pi^{vv(N)}_{vac}(q^2)$ and $\Pi^{tv(N)}_{vac}(q^2)$] is ultraviolet divergent
and dimensional regularization \cite{Hooft73,Peskin95,Cheng06} is used to isolate the
divergent parts. Since the mixing amplitude is generated by the difference between
the proton and neutron loop contributions, the divergent parts cancel out yielding
the vacuum amplitude finite.

\bea
\Pi^{\m\n}_{vac}(q^2) &=& \Pi^{\m\n(p)}_{vac}(q^2)-\Pi^{\m\n(n)}_{vac}(q^2)\nn \\
&=&\f{{\rm \grw}}{2\pi^2}q^2Q^{\m\n}\int^1_0 dx
\lt[(1-x)x + \f{C_\rho}{4} \rt] \nn \\
&\times& \ln\lt(\f{M^2_p -x(1-x)q^2}{M^2_n -x(1-x)q^2} \rt).
\label{pol:vac1}
\eea

Eq.(\ref{pol:vac1}) shows the four-momentum dependent vacuum polarization
tensor. From the above equation one can calculate longitudinal $(\P^L_{vac})$
and transverse $(\P^T_{vac})$ components of the vacuum mixing amplitude and
in the limit $q_0 \ra 0$, $\Pi^L_{vac}({\bf q}^2)= \Pi^T_{vac}({\bf q}^2)$.
Therefore, the average vacuum mixing amplitude is

\bea
\bar{\Pi}_{vac}({\bf q}^2) &=& \f{1}{3}\lt[\Pi^L_{vac}({\bf q}^2)
+ 2\Pi^T_{vac}({\bf q}^2) \rt] \nn \\
&=& - \f{{\rm \grw}}{12\pi^2}(2+3C_\rho)\ln\lt(\f{M_p}{M_n}\rt){\bf q}^2 \nn \\
& \equiv & - \mathcal{A}{\bf q}^2.
\label{pol:vac3}
\eea

Eq.(\ref{pol:vac3}) represents the three momentum dependent vacuum mixing amplitude.
This mixing amplitude vanishes for $M_n = M_p$ and then no CSV potential in vacuum will
exist.

To calculate density dependent mixing amplitude from  Eq.(\ref{pol:dens_vv0})
and (\ref{pol:dens_tv0}) we consider $E_N \approx M_N $. In the limit $q_0 \ra 0$,
one finds following expressions:

\bea
\Pi^{00(N)}_{med}({\bf q}^2) &=&-\f{{\rm \grw}}{4\pi^2 M_N}
\lt[\lt\{\f{4}{3}k^3_N - \f{1}{2}k_N{\bf q}^2 + 2k_NM^2_N \rt.\rt. \nn \\
&-&\lt.\lt. \lt(\f{{\bf q}^3}{8}-\f{{\bf q}k^2_N}{2}-\f{{\bf q}M^2_N}{2}
+ 2\f{M^2_Nk^2_N}{{\bf q}}\rt) \rt.\rt. \nn \\
&\times& \lt. \lt. \ln\lt(\f{{\bf q} - 2k_N}{{\bf q}+2k_N}\rt) \rt\}
+ \f{C_\rho}{2} \lt\{ {\bf q}^2k_N \rt. \rt. \nn \\
&+& \lt. \lt. \lt(\f{ {\bf q}^3}{4}-{\bf q}k^2_N\rt)
\ln\lt(\f{{\bf q}-2k_N} {{\bf q}+2k_N} \rt) \rt\} \rt]
\label{pol:dens_00} \\
&& \nn \\
\Pi^{11(N)}_{med}({\bf q}^2) &=& \f{{\rm \grw}}{4\pi^2M_N}
\lt[ \lt\{\f{1}{3}k^3_N -\f{3}{8}{\bf q}^2k_N \rt. \rt. \nn \\
&-& \lt.\lt. \lt(\f{3}{32}{\bf q}^3 + \f{k^4_N}{2{\bf q}}
+ \f{{\bf q}^2k_N}{4}\rt) \rt.\rt. \nn \\
&\times& \lt. \lt. \ln\lt(\f{{\bf q} - 2k_N}{{\bf q}+2k_N}\rt) \rt\}
+ \f{C_\rho}{2} \lt\{ {\bf q}^2k_N \rt. \rt. \nn \\
&+& \lt. \lt. \lt(\f{ {\bf q}^3}{4}-{\bf q}k^2_N\rt)
\ln\lt(\f{{\bf q}-2k_N} {{\bf q}+2k_N} \rt) \rt\} \rt]
\label{pol:dens_11}
\eea

Note that the terms within the first curly braces of both Eq.(\ref{pol:dens_00})
and Eq.(\ref{pol:dens_11}) arise from the vector-vector interaction while the terms
within the second curly braces arise from tensor-vector interaction of the density
dependent polarization tensor. The $33$ component of the density dependent polarization
tensor vanishes {\em i.e.} $\Pi^{33(N)}_{med}({\bf q}^2) = 0$. Now

\bea
\bar{\Pi}_{med}({\bf q}^2) = \f{1}{3}\lt[\Pi^L_{med}({\bf q}^2)
+ 2\Pi^T_{med}({\bf q}^2) \rt],
\eea

\noindent where

\bea
\Pi^L_{med}({\bf q}^2) &=& - \lt[\Pi^{00(p)}_{med}({\bf q}^2)
- \Pi^{00(n)}_{med}({\bf q}^2) \rt], \\
\Pi^T_{med}({\bf q}^2) &=& - \lt[\Pi^{11(p)}_{med}({\bf q}^2)
- \Pi^{11(n)}_{med}({\bf q}^2) \rt].
\eea

With the suitable expansion of Eqs.(\ref{pol:dens_00})
and (\ref{pol:dens_11}) in terms of $|{\bf q}|/k_{p(n)}$ and keeping
$\mathcal{O}({\bf q}^2/k^2_{p(n)})$ terms we get

\nointerlineskip
\bea
\bar{\Pi}_{med}({\bf q}^2) \simeq  \Delta^\prime - \mathcal{A}^\prime{\bf q}^2 ,
\label{pol:dens0}
\eea

\noindent where

\nointerlineskip
\bea
\Delta^\prime\!\! &=&\!\!\f{{\rm \grw}}{12\pi^2} \lt[ 3\lt(\f{k^3_p}{M_P}-\f{k^3_n}{M_n}\rt)
 + 4(k_pM_p - k_nM_n) \rt]. \label{dprm} \\
\mathcal{A}^\prime\!\! &=&\!\!\f{{\rm \grw}}{12\pi^2}\lt[3(1\!-\!C_\rho)\!
\lt(\f{k_p}{M_p}\!-\!\f{k_n}{M_n}\rt)\!+\!\f{1}{3}\lt(\f{M_p}{k_p}\!-\!\f{M_n}{k_n}\rt) \rt], \nn \\
&&  \label{aprm}
\eea

Clearly, $\bar{\Pi}_{med}({\bf q}^2)$ is also three momentum dependent and
if $M_n=M_p$ it vanishes in SNM but is non-vanishing in ANM. In the present
calculation nucleon masses are considered non-degenerate and the asymmetry
parameter $\a \neq 0$.

To construct CSV potential we take non-relativistic (NR) limit of the Dirac spinors
in which case we obtain

\bea
u_N(\vp) \simeq \lt(1-\f{{\bf P}^2}{8M^2_N}-\f{{\bf q}^2}{32M^2_N}\rt)
\left(\begin{array}{c}
1 \\
\f{\vs\cdot \lt({\bf P} + {\bf q}/2\rt)}{2M_{N}} \\
\end{array}
\right),
\label{sp}
\eea

\noindent where ${\bf \sigma}$ is the spin of nucleon. ${\bf P}$ denotes the
average three momentum of the interacting nucleon pair and ${\bf q}$  stands for
the three momentum of the meson.

The explicit expression of full CSV potential in momentum space
can be obtained from Eq.(13) of ref.\cite{Biswas08} by replacing
the mixing amplitude $\Pi_{\rho\o}({\bf q})$ with $\bar{\Pi}({\bf q}^2)$.

\bea
V^{NN}_{CSV}({\bf q})\!\!&=&\!-\f{{\rm g_\rho g_\o}~\bar{\Pi}({\bf q}^2)}
{({\bf q}^2 + m^2_\rho)({\bf q}^2 + m^2_\o)} \nn \\
&\times &\!\!\lt[T^+_3\lt\{\!\lt(\!1\!+\!\f{3{\bf P}^2}{2M^2_N}\!-\!\f{{\bf q}^2}{8M^2_N}
\!-\!\f{{\bf q}^2}{4M^2_N} ({\bf \s}_1 \cdot {\bf \s}_2) \rt. \rt. \rt. \nn \\
& + &\!\!\! \lt.\lt.\lt. \f{3i}{2M^2_N}{\bf S}\cdot ({\bf q}\times {\bf P})
\!+\!\f{1}{4M^2_N}({\bf \s}_1\cdot {\bf q})({\bf \s}_2\cdot {\bf q}) \rt.\rt. \rt. \nn \\
& + & \!\!\!\lt.\lt.\lt. \f{1}{M^2_N}({\bf \hat{q}}\cdot {\bf P})^2 \rt)
\!-\!\f{C_\rho}{2M} \lt(\f{{\bf q}^2}{2M_N}
\!+\! \f{{\bf q}^2}{2M_N}({\bf \s}_1 \cdot {\bf \s}_2) \rt.\rt. \rt. \nn \\
& - &\!\! \lt. \lt. \lt. \f{2i}{M_N}{\bf S} \cdot ({\bf q}\times {\bf P})
\!-\! \f{1}{2M_N}({\bf \s}_1 \cdot {\bf q})({\bf \s}_2\cdot {\bf q})\rt) \rt\} \rt. \nn \\
& - &\!\! \lt. T^-_3 \f{C_\rho}{2M} \lt\{\lt(\f{{\bf q}^2}{2M}
\!-\!\f{{\bf q}^2}{2M}({\bf \s}_1\cdot{\bf \s}_2) \rt. \rt. \rt. \nn \\
& + &\!\! \lt. \lt. \lt. \f{1}{2M}({\bf \s}_1\cdot{\bf q})({\bf \s}_2\cdot{\bf q})\rt)
\f{\Dt M(1,2)}{M} \rt. \rt. \nn \\
& - & \!\!\lt. \lt. \f{i}{M}({\bf \s}_1-{\bf \s}_2)\cdot({\bf q}\times{\bf P})\rt\}\rt].
\label{pot:mom_full}
\eea

Eq.(\ref{pot:mom_full}) presents the full CSV $NN$ potential in momentum
space in matter. Here $T^\pm_3 = \t_3(1)\pm\t_3(2)$ and ${\bf S}=\f{1}{2}({\bf\s}_1+{\bf\s}_2)$
is the total spin of the interacting nucleon pair. We define $M=(M_n+M_p)/2$,
$\Dt M=(M_n-M_p)/2$ and $\Dt M(1,2)=-\Dt M(2,1) = \Dt M $. It is be mentioned
that the spin dependent parts of the potential appear because of the contribution
of the external nucleon legs shown in Fig.\ref{fig00}. On the other hand, $3{\bf P}^2/2M^2_N$
and $-{\bf q}^2/8M^2_N$ arise due to expansion of the relativistic energy $E_N$ of the
spinors.

In matter, $V^{NN}_{CSV}({\bf q})$, consists of two parts, one contains
the vacuum mixing amplitude and other contains the density dependent mixing
amplitude. The former is denoted by $V^{NN}_{vac}({\bf q})$ and later by
$V^{NN}_{med}({\bf q})$. Thus, $V^{NN}_{CSV}({\bf q})= V^{NN}_{vac}({\bf q})
+ V^{NN}_{med}({\bf q})$.

From Eq.(\ref{pot:mom_full}) we extract the following term which, in coordinate
space gives rise to the $\delta$-function potential.

\bea
\delta V^{NN}_{CSV} &=& {\rm \grw}\lt(\mathcal{A}+\mathcal{A}^\prime\rt) \nn \\
&\times& \lt[ \lt(\f{1+2C_\rho}{8M^2_N}\rt) + \lt(\f{1+C_\rho}{4M^2_N}\rt)(\s_1\cdot\s_2) \rt].
\label{pot:contact}
\eea

To avoid the appearance of $\delta$-function potential in coordinate space, one should
introduce form factors [$F_i({\bf q}^2)$], for which the meson-nucleon coupling constants
become momentum dependent {\em i.e.} ${\rm g}_i({\bf q}^2) = {\rm g}_i F_i({\bf q}^2)$.
Here we use the following form  factor.

\bea
F_i({\bf q}^2) = \lt(\f{\L^2_i - m^2_i}{\L^2_i + {\bf q}^2} \rt),
\label{form}
\eea

\noindent where $\L_i$ is the cut-off parameter governing the range of the
suppression and $m_i$ denotes the mass of exchanged meson.

The full CSV potential presented in Eq.(\ref{pot:mom_full}) contains both Class $III$
and Class $IV$ potentials, and both break the charge symmetry in $NN$ interactions.
The terms within the first and the second curly braces represent Class $(III)$ and
Class $IV$ potentials, respectively. Class $(III)$ potential differentiates between
$nn$ and $pp$ systems while Class $(IV)$ $NN$ potential exists in the $np$ system only.
We, in this paper, restrict ourselves to Class $(III)$ potential only.

The coordinate space potential can be easily obtained by Fourier transformation of
$V^{NN}_{CSV}({\bf q})$ {\em i.e.}

\bea
V^{NN}_{CSV}({\bf r})=\int\f{d^3{\bf q}}{(2\pi)^3} V^{NN}_{CSV}({\bf q})~
e^{-i{\bf q\cdot r}}
\eea

We drop the term $3{\bf P}^2/2M^2_N$ from Eq.(\ref{pot:mom_full}) while taking
the Fourier transform as it is not important in the present context. However,
this term becomes important to fit $^1S_0$ and $^3P_2$ phase shifts simultaneously.

Now the CSV potential in coordinate space without $\delta V^{NN}_{CSV}$ reduces to

\bea
V^{NN}_{vac}({\bf r}) &=& -\f{\rm \grw}{4\pi} \mathcal{A}T^+_3
\lt[\lt(\f{m^3_\rho Y_0(x_\rho) - m^3_\o Y_0(x_\o)}{m^2_\o - m^2_\rho} \rt)\rt. \nn \\
&+& \f{1}{M^2_N}\lt.\lt(\f{m^5_\rho V_{vv}(x_\rho)-m^3_\o V_{vv}(x_\o)}{m^2_\o-m^2_\rho}\rt)\rt.\nn \\
&+& \f{C_\rho}{2M^2_N}\lt.\lt(\f{m^5_\rho V_{tv}(x_\rho)-m^3_\o V_{tv}(x_\o)}{m^2_\o-m^2_\rho}\rt)\rt],
\label{pot:vac_cord}
\eea

\noindent and

\bea
V^{NN}_{med}({\bf r})\!\!&=&\!\! -\f{\rm \grw}{4\pi} T^+_3
\lt[ \Delta^\prime \lt\{\lt(\f{m_\rho Y_0(x_\rho)-m_\o Y_0(x_\o)}
{m^2_\o-m^2_\rho} \rt) \rt. \rt. \nn \\
&+&\!\! \f{1}{M^2_N} \lt. \lt.\lt(\f{m^3_\rho V_{vv}(x_\rho)-m^3_\o V_{vv}(x_\o)}
{m^2_\o-m^2_\rho}\rt)\rt. \rt. \nn \\
&+&\!\! \f{C_\rho}{2M^2_N}\lt. \lt. \lt(\f{m^3_\rho V_{tv}(x_\rho)-m^3_\o V_{tv}(x_\o)}
{m^2_\o-m^2_\rho}\rt)\rt\} \rt. \nn \\
&+&\!\!\mathcal{A}^\prime\!\! \lt. \lt\{\lt(\f{m^3_\rho Y_0(x_\rho)-m^3_\o Y_0(x_\o)}
{m^2_\o-m^2_\rho} \rt)\rt. \rt. \nn \\
&+&\!\!\f{1}{M^2_N} \lt. \lt.\lt(\f{m^5_\rho V_{vv}(x_\rho)-m^5_\o V_{vv}(x_\o)}
{m^2_\o-m^2_\rho}\rt)\rt. \rt. \nn \\
&+&\!\!\f{C_\rho}{2M^2_N}\lt. \lt. \lt(\f{m^5_\rho V_{tv}(x_\rho)-m^5_\o V_{tv}(x_\o)}
{m^2_\o-m^2_\rho}\rt)\rt\} \rt],
\label{pot:dens_cord}
\eea

\noindent where $x_i = m_i r$. The explicit expressions of $V_{vv}(x_i)$ and
$V_{tv}(x_i)$ are given in Ref.\cite{Biswas08}. In Eqs.(\ref{pot:vac_cord}) and
(\ref{pot:dens_cord}), the $M^{-2}_N$ independent terms represent central parts
without contributions of external nucleon legs.

The potentials presented in Eqs.(\ref{pot:vac_cord}) and (\ref{pot:dens_cord})
do not include the form factors so that these potentials diverge near the core.
The problem of divergence near the core can be removed by incorporating form
factors as discussed before. With the inclusion of form factors, $V^{NN}_{vac}({\bf r})$
and $V^{NN}_{med}({\bf r})$ take the following form:

\bea
V^{NN}_{vac}({\bf r})\!\! &=&\!\! -\f{{\rm \grw}}{4\pi} \mathcal{A}T^+_3
\! \lt[ \lt(\!\f{a_\rho m^3_\rho Y_0(x_\rho)-a_\o m^3_\o Y_0(x_\o)}
{m^2_\o-m^2_\rho}\!\rt)\rt. \nn \\
&+&\!\! \lt.\f{1}{M^2_N}\!\lt(\!\f{a_\rho m^5_\rho V_{vv}(x_\rho)\!-\!a_\o m^5_\o V_{vv}(x_\o)}
{m^2_\o-m^2_\rho}\!\rt)\rt. \nn \\
&+&\!\! \lt.\f{C_\rho}{2M^2_N}\!\lt(\!\f{a_\rho m^5_\rho V_{tv}(x_\rho)\!-\!a_\o m^5_\o V_{tv}(x_\o)}
{m^2_\o-m^2_\rho}\!\rt)\rt. \nn \\
&-&\!\!\lambda\! \lt. \lt\{\!\lt(\f{b_\rho \L^3_\rho Y_0(X_\rho)\!-\!b_\o \L^3_\o Y_0(X_\o)}
{m^2_\o-m^2_\rho}\!\rt)\rt. \rt. \nn \\
&+&\!\!\! \lt.\lt. \f{1}{M^2_N}\!\lt(\!\f{b_\rho \L^5_\rho V_{vv}(X_\rho)\!-\!b_\o \L^5_\o V_{vv}(X_\o)}
{m^2_\o-m^2_\rho}\!\rt)\rt. \rt. \nn \\
&+&\!\!\!\lt.\lt. \f{C_\rho}{2M^2_N}\!\lt(\!\f{b_\rho \L^5_\rho V_{tv}(X_\rho)\!-\!b_\o \L^5_\o V_{tv}(X_\o)}{m^2_\o-m^2_\rho}\!\rt)\!\!\rt\}\!\rt],
\label{pot:vac_ff}
\eea
\noindent and
\bea
V^{NN}_{med}({\bf r})\!\! &=&\!\!-\f{{\rm \grw}}{4\pi}T^+_3
\!\!\lt[ \Delta^\prime\!\lt\{\!\!\lt(\!\f{a_\rho m_\rho Y_0(x_\rho)-a_\o m_\o Y_0(x_\o)}
{m^2_\o-m^2_\rho}\!\rt)\rt.\rt. \nn \\
&+& \!\!\!\lt. \lt.\f{1}{M^2_N}\!\lt(\!\f{a_\rho m^3_\rho V_{vv}(x_\rho)\!-\!a_\o m^3_\o V_{vv}(x_\o)}
{m^2_\o-m^2_\rho}\!\rt)\rt.\rt. \nn \\
&+&\!\!\!\lt. \lt.\f{C_\rho}{2M^2_N}\!\lt(\!\f{a_\rho m^3_\rho V_{tv}(x_\rho)\!-\!a_\o m^3_\o V_{tv}(x_\o)}
{m^2_\o-m^2_\rho}\!\rt)\!\rt\}\rt. \nn \\
&+&\!\!\!\mathcal{A}^\prime\!\!\lt.\lt\{\!\lt(\!\f{a_\rho m^3_\rho Y_0(x_\rho)\!-\!a^3_\o m_\o Y_0(x_\o)}
{m^2_\o-m^2_\rho}\!\rt)\rt.\rt. \nn \\
&+&\!\!\!\lt.\lt.\f{1}{M^2_N}\!\lt(\!\f{a_\rho m^5_\rho V_{vv}(x_\rho)\!-\!a_\o m^5_\o V_{vv}(x_\o)}
{m^2_\o-m^2_\rho}\!\rt)\rt.\rt. \nn \\
&+&\!\!\!\lt.\lt.\f{C_\rho}{2M^2_N}\!\lt(\!\f{a_\rho m^5_\rho V_{tv}(x_\rho)\!-\!a_\o m^5_\o V_{tv}(x_\o)}
{m^2_\o-m^2_\rho}\!\rt)\!\rt\}\rt. \nn \\
&-&\!\!\lambda \Delta^\prime \!\lt.\lt\{\!\lt(\!\f{b_\rho \L_\rho Y_0(X_\rho)\!-\!b_\o \L_\o Y_0(X_\o)}
{m^2_\o-m^2_\rho}\!\rt)\rt. \rt. \nn \\
&+&\!\! \lt.\lt. \f{1}{M^2_N}\!\lt(\!\f{b_\rho \L^3_\rho V_{vv}(X_\rho)\!-\!b_\o \L^3_\o V_{vv}(X_\o)}
{m^2_\o-m^2_\rho}\!\rt)\rt. \rt. \nn \\
&+&\!\!\!\lt.\lt. \f{C_\rho}{2M^2_N}\!\lt(\!\f{b_\rho \L^3_\rho V_{tv}(X_\rho)\!-\!b_\o \L^3_\o V_{tv}(X_\o)}{m^2_\o-m^2_\rho}\!\rt)\!\rt\}  \rt. \nn\\
&-&\!\!\lambda \mathcal{A}^\prime\!\!\lt.\lt\{\!\lt(\f{b_\rho \L^3_\rho Y_0(X_\rho)\!-\!b_\o \L^3_\o Y_0(X_\o)}{m^2_\o-m^2_\rho}\!\rt)\rt. \rt. \nn \\
&+&\!\!\! \lt.\lt. \f{1}{M^2_N}\!\lt(\!\f{b_\rho \L^5_\rho V_{vv}(X_\rho)\!-\!b_\o \L^5_\o V_{vv}(X_\o)}
{m^2_\o-m^2_\rho}\!\rt)\rt. \rt. \nn \\
&+&\!\!\!\lt.\lt.\f{C_\rho}{2M^2_N}\!\lt(\!\f{b_\rho \L^5_\rho V_{tv}(X_\rho)\!-\!b_\o \L^5_\o V_{tv}(X_\o)}{m^2_\o-m^2_\rho}\!\rt) \rt\}\!\rt].
\label{pot:dens_ff}
\eea

In Eqs.(\ref{pot:vac_ff}) and (\ref{pot:dens_ff}), $X_i = \L_i r$ and

\bea
\lambda &=& \lt(\f{m^2_\o - m^2_\rho}{\L^2_\o - \L^2_\rho}\rt), \label{lambda} \\
a_i = \lt(\f{\L^2_j - m^2_j}{\L^2_j - m^2_i}\rt),&&
b_i = \lt(\f{\L^2_j - m^2_j}{m^2_j - \L^2_i}\rt), \label{ab} \\
{\rm where}~~i(j) = \rho, \o &&(i \neq j). \nn
\eea

Note that Eqs.(\ref{pot:vac_ff}) and (\ref{pot:dens_ff}) contain the contribution
of $\delta V^{NN}_{CSV}$ and the problem of divergence near the core is removed.

\section{results}

\begin{figure}[hb!]
\begin{center}
\resizebox{6.5cm}{5cm}{\includegraphics[]{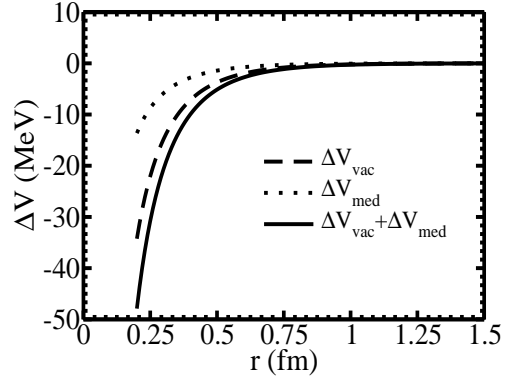}}
\caption{$\Dt V$ for $^1$S$_0$ state (without $\dt V^{NN}_{CSV}$ and form factors)
at $\rho_B = 0.148$ fm$^{-3}$ and $\a = 1/3$.
\label{fig01}}
\end{center}
\end{figure}

Using Eqs.(\ref{pot:vac_cord}) and (\ref{pot:dens_cord}) we show the difference
of CSV potentials between $nn$ and $pp$ systems {\em i.e.} $\Dt V=V^{nn}_{CSV}-V^{pp}_{CSV}$
in Fig.\ref{fig01} for $^1$S$_0$ state at nuclear matter density $\rho_B = 0.148$
fm$^{-3}$ with asymmetry parameter $\a = 1/3$. The dashed and dotted curves show
$\Dt V_{vac}$ and $\Dt V_{med}$, respectively. The total contribution ({\em i.e.}
$\Dt V_{vac} + \Dt V_{med}$) is shown by the solid curve. It is observed that the
density dependent CSV potential can not be neglected while estimating CSV observables
such as binding energy difference of mirror nuclei.\\

\begin{figure}[htb!]
\begin{center}
\resizebox{6.5cm}{5cm}{\includegraphics[]{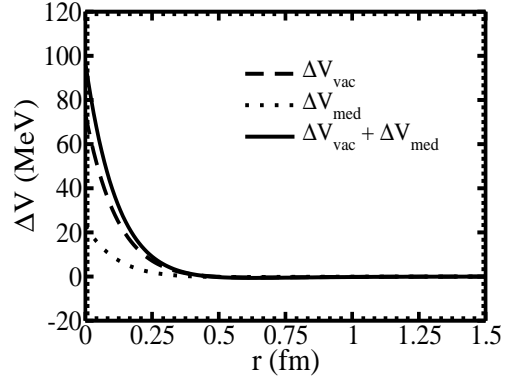}}
\caption{$\Dt V$ for (with form factors including the Fourier transform of
$\dt V^{NN}_{CSV}$) $^1$S$_0$ state at $\rho_B = 0.148$ fm$^{-3}$
and $\a = 1/3$.
\label{fig02}}
\end{center}
\end{figure}

Fig.\ref{fig02} displays $\Dt V$ for $^1$S$_0$ state including the Fourier
transform of $\dt V^{NN}_{CSV}$ and form factors. $\Dt V_{vac}$ and $\Dt V_{med}$
are represented by dashed and dotted curves. Solid curve shows the sum of these
two contributions. Note that incorporation of form factors remove the problem
of divergence near the core. It is observed that the CSV $NN$ potential changes
its sign due to inclusion of the Fourier transform of $\dt V^{NN}_{CSV}$.

\section{summary and discussion}

In this work we have constructed the CSV $NN$ potential in dense matter
using asymmetry driven momentum dependent $\rho^0$-$\omega$ mixing amplitude
within the framework of one-boson exchange model. Furthermore, the correction
to the central part of the CSV potential due to external nucleon legs are also
considered. The closed-form analytic expressions both for vacuum
and density dependent CSV $NN$ potentials in coordinate space are presented.

We have shown that the vacuum mixing amplitude and the density dependent mixing
amplitude are of similar order of magnitude and both contribute with the same sign
to the CSV potential. The contribution of density dependent CSV potential is not
negligible in comparison to the vacuum CSV potential.

\appendix
\section{}

The position space Fermion propagator in vacuum is given by the vacuum expectation
value of the time ordered product of Fermion fields.

\bea
i\tilde{G}_N(x-x^\prime)&=& \bra 0|\mathcal{T}[\psi(x)\bar{\psi}(\xp)]|0\ket. \label{app01}
\eea

In medium, the vacuum $|0\ket$ is replaced by the ground state $|\Psi_0\ket$ which
contains positive-energy particles with same Fermi momentum $k_N$ and no antiparticles.
Thus,

\bea
i\tilde{G}_N(x-x^\prime)&=& \bra \Psi_0|\psi(x)\bar{\psi}(\xp)|\Psi_0\ket \th(t-\tp) \nn \\
&-& \bra \Psi_0|\bar{\psi}(\xp)\psi(x)|\Psi_0\ket \th(\tp-t). \label{app02}
\eea

Note that the time-ordered product in Eq.(\ref{app02}) involves negative sign for Fermions.
The Fermion field contains both the positive- and negative-energy solutions. The modal
expansion for the Fermion fields are,

\bea
\psi(x)\!\!&=&\!\!\!\!\int\!\!\f{d^3\vk}{\sq{(2\pi)^32E_k}}\!\sum_s\!
\lt(\!a_{ks}u_{ks}e^{-ik\cdot x}\!\!+\!b^\dagger_{ks}v_{ks}e^{ik\cdot x}\!\rt), \nn \\
&& \label{app03}\\
\bar{\psi}(x)\!\!&=&\!\!\!\!\int\!\!\f{d^3\vk}{\sq{(2\pi)^32E_k}}\!\sum_s\!
\lt(\!a^\dagger_{ks}\bar{u}_{ks}e^{ik\cdot x}\!\!+\!b_{ks}\bar{v}_{ks}e^{-ik\cdot x}\!\rt). \nn \\
&& \label{app04}
\eea

Here $a^\dagger_{ks}$ and $a_{ks}$ are the creation and annihilation operators
for particles and likewise $b^\dagger_{ks}$ and $b_{ks}$ are the creation
and annihilation operators for antiparticles. The only nonvanishing anticommutation
relations are

\bea
\{a_{ks},a^\dagger_{\kp\sp}\} = \{b_{ks},b^\dagger_{\kp\sp}\} = \dt_{s\sp}\dt^3(\vk-\vkp).
\label{app05}
\eea

Since $|\Psi_0\ket$ contains only positive-energy particles, we have the following
relations:

\be
\lt.
\begin{array}{ccl}
b_{ks}|\Psi_0\ket & = 0 & {\rm for~all}~\vk \\
&&\\
a_{ks}|\Psi_0\ket & = 0 & {\rm for}~|\vk|>k_N \\
&&\\
a^\dagger_{ks}|\Psi_0\ket & = 0 & {\rm for}~|\vk|<k_N \\
&&\\
a_{ks}a^\dagger_{ks}|\Psi_0\ket&= n({\bf k})|\Psi_0\ket & \\
\end{array}
\rt.
\label{app06}
\ee

$n({\bf k})$ is either $0$ or $1$ and this can be accomplished with the step
function $\th(k_N-|{\bf k}|)$. Using Eqs.(\ref{app03})-(\ref{app06}) one obtains

\bea
&&\bra\Psi_0|\psi(x)\bar{\psi}(\xp)|\Psi_0\ket\!=\!\int\!\f{d^3\vk}{\sq{(2\pi)^32E_k}}
\!\int\!\f{d^3\vkp}{\sq{(2\pi)^32E_{\kp}}} \nn \\
&\times&\!\!\sum_{s\sp} \!\bra\Psi_0|a_{ks}a^\dagger_{\kp\sp}|\Psi_0\ket u_{ks}\bar{u}_{\kp\sp}
e^{-i(k\cdot x - \kp\cdot\xp)} \nn \\
&=&\!\!\!\int\!\f{d^3\vk}{(2\pi)^32E_k}(k\!\!\!/+M_N) e^{-ik\cdot(x\!-\!\xp)}\![\!1\!-\!\th(k_N\!-\!|\vk|)\!], \label{app07}
\eea
\noindent and
\bea
&&\bra\Psi_0|\bar{\psi}(\xp)\psi(x)|\Psi_0\ket\!=\!\!\int\!\f{d^3\vk}{\sq{(2\pi)^32E_k}}
\!\int\!\f{d^3\vkp}{\sq{(2\pi)^32E_{\kp}}} \nn \\
&\times&\!\!\sum_{s\sp}\!\lt[\!\bra\Psi_0|a^\dagger_{\kp\sp}a_{ks}|\Psi_0\ket\bar{u}_{\kp\sp}u_{ks}
e^{-i(k\cdot x\! -\!\kp\cdot\xp)}\rt.\nn\\
&+&\lt.\!\! \bra\Psi_0|b_{\kp\sp}b^\dagger_{ks}|\Psi_0\ket \bar{v}_{\kp\sp}v_{ks}
e^{i(k\cdot x\!-\!\kp\cdot\xp)}\!\rt] \nn \\
&=&\!\!\!\int\!\f{d^3\vk}{(2\pi)^32E_k} \lt[\!(k\!\!\!/+M_N)e^{-ik\cdot(x\!-\!\xp)}\th(k_N\!-\!|\vk|)
\rt.\nn\\
&+&\!\!\!\lt. \!(k\!\!\!/-M_N)e^{ik\cdot(x\!-\!\xp)}\!\rt].\label{app08}
\eea

\noindent Now,

\bea
\th(t-\tp)e^{-ik\cdot(x-\xp)}\!\!&=&\!\!i\int\f{dk_0}{2\pi}\f{e^{-ik\cdot(x-\xp)}}{k_0-E_k+i\ep}. \label{app09}\\
\th(\tp\!-\!t)e^{-ik\cdot(x-\xp)}\!\! &=&\!\!-i\!\int\f{dk_0}{2\pi}\f{e^{-ik\cdot(x-\xp)}}{k_0\!-\!E_k\!-\!i\ep},
\label{app10}\\
\th(\tp\!-\!t)e^{ik\cdot(x-\xp)}\!\!&=&\!\!i
\!\int\f{dk_0}{2\pi}\f{e^{ik\cdot(x-x)}}{k_0\!-\!E_k\!+\!i\ep}.
\label{app11}
\eea

From Eqs.(\ref{app08}), (\ref{app10}) and (\ref{app11}) we have

\bea
&& \bra\Psi_0|\bar{\psi}(\xp)\psi(x)|\Psi_0\ket \th(t-\tp) \nn \\
&=&\!\!\!-i\!\int\!\f{d^4k}{(2\pi)^42E_k}e^{-ik\cdot(x-\xp)}\!(k\!\!\!/+M_N)\f{\th(k_N-|\vk|)}{k_0-E_k-i\ep} \nn \\
&+&\!\!i\int\!\f{d^4k}{(2\pi)^42E_k}(k\!\!\!/-M_N)\!\f{e^{ik\cdot(x-\xp)}}{k_0-E_k+i\ep}.
\label{app12}
\eea

Now changing $k\ra -k$ in the last integral of Eq.(\ref{app12}) and substituting
Eqs.(\ref{app07}), (\ref{app09}) and (\ref{app12}) in Eq.(\ref{app02}) we get,

\bea
&&i\tilde{G}_N(x-\xp)=\! i\!\int\! \f{d^4k}{(2\pi)^42E_N}e^{-ik\cdot(x-\xp)}(k\!\!\!/\!+\!M_N)~~~\nn \\
&\times&\!\!\!\lt[\!\f{1\!-\!\th(k_N\!-\!|\vk|)}{k_0\!-\!E_N\!+\!i\ep}\!+\!\f{\th(k_N\!-\!|\vk|)}
{k_0\!-\!E_N\!-\!i\ep}\!-\!\f{1}{k_0\!+\!E_N\!-\!i\ep}\!\rt]. \label{app13}
\eea

In Eq.(\ref{app13}) $E_k$ has been replaced by $E_N$. The first term of Eq.(\ref{app13})
represents particle propagation above the Fermi sea and the second term indicates
the propagation of holes inside the Fermi sea. The last term shows the propagation
of holes in the infinite Dirac sea. Now,

\bea
\f{1}{k_0\!-\!E_N\!+\!i\ep}\!-\!\f{1}{k_0\!+\!E_N\!-\!i\ep}\!\!&=&\!\!\f{2E_N}{k^2\!-\!M^2_N\!+\!i\z},
\label{app14} \\
\f{1}{k_0\!-\!E_N\!-\!i\ep}\!-\!\f{1}{k_0\!-\!E_N\!+\!i\ep}\!\!&=&\!\!2i\pi\dt(k_0\!-\!E_N).\label{app15}
\eea

From Eqs.(\ref{app13})-(\ref{app15}),

\bea
i\tilde{G}_N(x-\xp)\!&=&\!\! i\int\!\f{d^4k}{(2\pi)^4}e^{-ik\cdot(x-\xp)} G_N(k) \label{app16}
\eea

where $G_N(k)$ is the in-medium Fermion propagator in momentum space.

\bea
G_N(k)\!\!&=&\!\!\f{k\!\!\!/\!+\!M_N}{k^2\!-\!M^2_N\!+\!i\ep} \nn \\
&+&\!\!\f{i\pi}{E_N}(k\!\!\!/\!+\!M_N)\dt(k_0\!-\!E_N)\th(k_N\!-\!|\vk|) \nn \\
&=& G_N^F(k)+G_N^D(k). \label{app17}
\eea

\end{document}